\documentclass{article}

\usepackage{geometry}
\usepackage{amsmath}
\usepackage{amssymb}
\usepackage{units}

\usepackage{particles}
\usepackage{numinsec}
\usepackage{pptitle}

\title{Extended Electroweak Interactions and the Muon $g_\muon-2$}
\author{Kevin R. Lynch\thanks{krlynch@bu.edu}\\ \\
Department of Physics\\ Boston University\\ 590 Commonwealth Avenue\\
Boston, MA 02215}
\date{August 7, 2001}
\preprintwidth{hep-ph/0108080}
\preprints{BUHEP-01-16\\hep-ph/0108080}

\newcommand{\SMgauge}{\ensuremath{{\SUtwo_L \times \Uone_Y}}}
\newcommand{\Uoneem}{\ensuremath{{\Uone_{\text{em}}}}}

\newcommand{\Grm}{\ensuremath{{\mathrm{G}}}}
\newcommand{\geff}{\ensuremath{{g_\effective}}}

\newcommand{\modelA}{\ensuremath{{\SUtwo_1 \times \SUtwo_2 \times
    \Uone_Y}}}
\newcommand{\modelB}{\ensuremath{{\SUtwo_L \times \SUtwo_2 \times
    \Uone_3}}}
\newcommand{\modelC}{\ensuremath{{\SUtwo_L \times \Uone_1 \times
    \Uone_2}}}

\newcommand{\amuon}{\ensuremath{{a_\muon}}}

\newcommand{\sphi}{\ensuremath{{s_\phi}}}
\newcommand{\cphi}{\ensuremath{{c_\phi}}}
\newcommand{\tphi}{\ensuremath{{t_\phi}}}

\newcommand{\sW}{\ensuremath{{s_W}}}

\newcommand{\Ztilde}{{\ensuremath{{\particle{\tilde{Z}}}}}}
\newcommand{\Zboson}{{\ensuremath{{\particle{Z}}}}}
\newcommand{\Zright}{{\ensuremath{{\particle{Z_R}}}}}
\newcommand{\Wright}{{\ensuremath{{\particle{W_R}}}}}

\newcommand{\PsiG}{{\ensuremath{{\particle{\Psi}_G}}}}
\newcommand{\PsiGbar}{{\ensuremath{{\DiracBar{\particle{\Psi}}_G}}}}
\newcommand{\PsiM}{{\ensuremath{{\particle{\Psi}_M}}}}
\newcommand{\PsiMbar}{{\ensuremath{{\DiracBar{\particle{\Psi}}_M}}}}

\newcommand{\internal}{\ensuremath{{\text{int}}}}
\newcommand{\effective}{\ensuremath{{\text{eff}}}}

\newcommand{\dd}[2][]{\ensuremath{{\mathrm{d}^{#1}{#2}}}}

\begin{document}

\begin{titlepage}
\maketitle

\begin{abstract}  
  We look at a number of simple, but representative, models of
  extended electroweak gauge structures, and present the general
  contributions to \amuon\ from the heavy \Zprime\ and \Wprime\ 
  electroweak gauge bosons.  Of the models we have examined, none can
  explain the observed discrepancy between the current experimental
  value of \amuon\ and the Standard Model prediction if we require
  that the gauge fields explain the discrepancy by themselves.  In the
  context of models with new matter fields as well as the additional
  gauge fields discussed here, however, the gauge field contributions
  to \amuon\ can be a substantial and important part of the
  discrepancy.  
\end{abstract}

\end{titlepage}

\section{Introduction}
\label{sec:intro}

The recent measurement of the muon anomalous magnetic moment, \amuon,
by the Brookhaven E821 Collaboration \cite{Brown:2001mg} has raised
the tantalizing possibility that new physics lies within the reach of
current (or soon to be conducted) collider experiments.  If we assume that the
discrepancy between this measurement and the Standard Model
prediction,
\begin{equation}
\delta\amuon = \amuon(\text{exp}) - \amuon(\text{SM}) = 426(165)
\times 10^{-11}\ , 
\end{equation}
really is due to new physics, we should consider all possible
mechanisms that may generate a value of this size and sign.  Many
authors have weighed in with possible explanations of the discrepancy,
including supersymmetric scenarios \cite{Everett:2001tq,Feng:2001tr},
muon substructure \cite{Lane:2001ta}, leptoquark models
\cite{Mahanta:2001yc,Chakraverty:2001yg,Cheung:2001ip}, scenarios with
extra dimensions \cite{Agashe:2001ra,Appelquist:2001jz}, and exotic
fermions \cite{Choudhury:2001ad}.

In this letter, we would like to consider a different class of models,
namely ``pure gauge extensions'' to the \SMgauge\ Standard Model
electroweak gauge structure.  By ``pure gauge extension'' we mean that
we add additional gauge groups, those scalars with non-zero VEVs that
are necessary to break the gauge symmetries, and those spectator
fermions necessary to cancel gauge anomalies, but no other degrees of
freedom.  We are motivated to consider these models by the fact that
the difference between experiment and the Standard Model prediction is
of the same order, and has the same sign, as the Standard Model weak
contribution: to two loop order, the Standard Model weak contribution
is $\amuon(\text{weak}) = 152(4) \times 10^{-11}$, with the three loop
contribution predicted to be negligible compared to this value (for a
recent review to the theoretical state of the art, we cite
\cite{Czarnecki:2001pv} and the references therein).  In this letter,
we will consider only the one loop contributions of the extended gauge
symmetries; extrapolating from the Standard Model, we might expect the
two-loop expressions to reduce the contributions found here by a few
percent, but the precise size of the contributions is less crucial to
our explorations than are the general results we obtain.

Other authors have considered contributions to \amuon\ from new gauge
bosons \cite{Choudhury:2001ad,Huang:2001zx}. We will consider a
different class of models, based on the extended gauge groups $\SUtwo
\times \SUtwo \times \Uone$ and $\SUtwo \times \Uone \times \Uone$.
First, we will discuss the general structure and properties of models
of these types, and present relations which can be used to calculate
the contributions of general electroweak \Zprime\ and \Wprime\ bosons
to the muon anomaly.  We will then apply these results to three
specific classes of extended electroweak models: lepton-quark
non-universal models (the ununified models), left-right symmetric
models, and generation non-universal models.\footnote{While most of
  the explicit models we consider are motivated by dynamical symmetry
  breaking, similar gauge structures arise from different motivations,
  in particular from models motivated by string theory and grand
  unification; for an overview, we direct the reader to
  \cite{Langacker:2000ju} and the references therein.}  In the most
general case of models with new gauge bosons, the contributions to
\amuon\ can be arbitrarily large. In the class of models we consider
here, the extra gauge bosons decouple from the theory as their masses
increase.  Since we have not seen these gauge bosons directly or
indirectly in the electroweak data, we would not generally expect to
see contributions to \amuon as large as $\delta\amuon$.  However,
there are a number of reasons we should consider the magnitude of
contributions in these models.  First, both the errors on the data and
the standard model theoretical contributions are large compared to
both the Standard Model electroweak contributions and the measured
discrepancy and the final value may be considerably smaller than the
current value.  Second, in the context of these models, it is
important to determine what fraction of the discrepancy can be
accounted for by the new gauge physics, in order to determine what
other types of new physics, such as new fermions or scalars, might be
necessary to describe all of the available data in the context of
extended electroweak gauge models.  In addition to these decoupling
scenarios, we will consider the effects of fermion mixing in the
generation non-universal models, where the new gauge structure admits
tree level, flavor changing couplings in the charged lepton sector
which do not generically decouple.  In all cases, we will use the
measured muon anomaly to either constrain the masses of the new gauge
fields, under the assumption that the new fields are responsible for
all of the discrepancy, or we will use precision electroweak bounds on
the masses of the new fields to find the maximum contribution they can
make to the discrepancy.  We will close this paper by drawing some
general conclusions and will suggest future directions in model
building in light of our results.

\section{General Results}
\label{sec:general}

Any extended electroweak gauge model must have the experimentally well
verified Standard Model \SMgauge\ gauge structure at low energy.
Despite the strong experimental constraints on the properties of the
electroweak sector, there are still numerous gauge extensions that can
both satisfy the constraints and permit interesting, relatively
low-energy (that is, sub-TeV) phenomenology.  We consider the general
properties of the following electroweak gauge extensions:
\begin{gather*}
\modelA \stackrel{u}{\longrightarrow} \SMgauge
\stackrel{v}{\longrightarrow} \Uoneem\\   
\modelB \stackrel{u}{\longrightarrow} \SMgauge
\stackrel{v}{\longrightarrow} \Uoneem\\ 
\modelC \stackrel{u}{\longrightarrow} \SMgauge
\stackrel{v}{\longrightarrow} \Uoneem\ . 
\end{gather*}
In each case, the extended symmetry is broken by the vacuum
expectation value (VEV) of some scalar object $\Sigma$ (which may be
fundamental or composite) at energy scale $u$, followed by the
Standard Model breakdown by $\Phi$ at energy scale $v$.  We assume
that the higher breaking scale, $u$, is large compared to the Standard
Model breaking scale $v$ (such that $u^2/v^2 > 1$); this ensures
that the new contributions to \amuon\ will be dominated by the
additional heavy gauge fields and not the small shifts in the
couplings of the Standard Model gauge fields.  Current limits from
precision electroweak data ensure that this is true in the specific
models we examine later.

In order to review the structure of the couplings that arise in the
models above, we generalize our notation for the groups
\begin{equation*}
\Grm_1 \times \Grm_2 \times \Grm_3 \to \SMgauge \to
\Uoneem\ ,
\end{equation*}
which gives rise to the covariant derivative (displaying for
simplicity only the neutral sector)
\begin{equation*}
D_\mu = \partial_\mu - i \sum_{i=1}^3 g_i \particle{A}_\mu^i \tau_i \ ,
\end{equation*}
with diagonal generators $\tau_i$.  At the scale $u$, a first stage of
breaking occurs. Two of these groups (take $\Grm_2$ and $\Grm_3$) mix,
leaving a diagonal, unbroken group $\Grm_{2'}$, and a massive gauge
field, \Ztilde\ which couples approximately to
\begin{equation}
\tilde{g} \left(\frac{\cos\phi}{\sin\phi} \tau_3 - \frac{\sin\phi}{\cos\phi}
  \tau_2 \right)\ , 
\end{equation}
where $\phi$ is the mixing angle between the unbroken and broken gauge
fields, and $\tilde{g} = g_3 \sin\phi = g_2 \cos\phi$.  At the scale
$v$, a second stage of breaking occurs, and the remaining two unbroken
groups ($\Grm_1$ and $\Grm_{2'}$) mix, leaving an unbroken \Uoneem\ 
and a second massive (but lighter) gauge boson, \Zboson.  The remaining,
unbroken gauge group gives rise to an exactly massless photon, with
generator
\begin{equation}
e \left(\tau_1 + \tau_2 + \tau_3\right)\ ,
\end{equation}
while the generator of the second massive gauge field is
\begin{equation}
\frac{g}{\cos\thetaW} \left(T^3_L - \sin^2\thetaW Q\right)\ .
\end{equation}
The generators of the \Znaught\ and \Zprime\ mass eigenstates differ
from those above (the \Zboson\ and \Ztilde) by order one terms,
multiplied by powers of $v^2/u^2$; the differences are negligibly
small to the order we are working, and we will not consider them here.
Any charged gauge fields in the model obtain similar generators, but
these are more model dependent and will not be discussed in detail
here.

We note in passing that even larger gauge extensions, such as $\SUtwo
\times \SUtwo \times \Uone \times \Uone$, are possible; see, for
example, \cite{Rajpoot:1990hi}.  We will not consider them explicitly
here because, in general, we expect that the lowest mass vectors will
posses many of the properties of the vectors we study here, and the
additional heavier states will make negligible contributions to
\amuon.  In general, we expect that our general results will be
independent of the precise details of such large extensions.

The gauge structure of these models assures that, to lowest order, the
couplings of the new gauge fields to the photon will have the same
structure as for the Standard Model electroweak gauge fields.  In
particular, there will be no new multi-gauge boson vertices such as
$\Zprime\Znaught\photon$ or $\Wprime\Wpart\photon$.  With this
restriction, we can find the general one-loop contributions of new
charged and neutral vectors to \amuon.

The one-loop contribution in Feynman gauge from charged vectors in the
narrow width approximation ($\Gamma_\Wprime = 0$) is given by
\begin{multline}
\amuon^\Wprime = \frac{m_\muon \geff^2}{4 \pi^2} \int\limits_0^1
\dd{u} u^2 \frac{m_\muon (2u+1) 
  \left(vv^\dagger+aa^\dagger\right) - 3m_\internal
  \left(vv^\dagger-aa^\dagger\right)}{(1-u)m_\internal^2 + uM^2 +
  u(u-1) m_\muon^2}\\
- \frac{m_\muon}{8\pi^2} \int\limits_0^1 \dd{u}
\frac{u^2\left((m_\internal-m_\muon) vv^\dagger
    -(m_\internal+m_\muon) aa^\dagger\right) }{uM^2
  +(1-u)m_\internal^2 +u(u-1)m_\muon^2}\\
-\frac{m_\muon \geff^2}{8\pi^2} \frac{1}{M^2}
\int\limits_0^1 \dd{u} u(1-u) \left\{\frac{m_\muon\left\{
    (m_\internal-m_\muon)^2 vv^\dagger
    +(m_\internal+m_\muon)^2 aa^\dagger \right\}
  u}{uM^2+(1-u)m_\internal^2 +u(u-1)m_\muon^2}\ + \right. \\ 
\left. \frac{m_\internal \left\{(m_\internal-m_\muon)^2 vv^\dagger
      - (m_\internal+m_\muon)^2
      aa^\dagger\right\}}{uM^2+(1-u)m_\internal^2
    +u(u-1)m_\muon^2} \right\}
\ ,
\end{multline}
where the first line comes from a diagram with two \Wprime\ bosons in
the loop, the second line contains the contributions of the two
diagrams where one vector is replaced by the unphysical scalar, and
the final two lines are the contribution where both vectors are
replaced by unphysical scalars.  Most of the terms in the above
expressions arise from our definition of the coupling between the
gauge fields, the muon, and a neutrino\footnote{Neutrino here refers
  to any neutral fermion in the extended model with appropriate
  quantum numbers to couple to the muon.} in the Lagrangian
\begin{equation*}
\mathcal{L} \sim \sum_i \geff \DB{\muon} \DG{\mu} \left(v^i + a^i
  \DG{5}\right) \neutrino_i \Wprime_\mu + \text{h.c.} = \sum_i \geff
\DB{\muon}_L \DG{\nu} C_L^i \neutrino_{Li} 
\Wprime_\nu + \geff \DB{\muon}_R \DG{\nu} C_R^i \neutrino_{Ri}
\Wprime_\nu + \text{h.c.}\ ,
\end{equation*}
We have written the interaction both in terms of the vector, $v^i$,
and axial, $a^i$, couplings and in terms of the left-, $C_L^i$, and
right-handed, $C_R^i$, chiral couplings,\footnote{The different types
  of couplings are related by
\begin{equation*}
v = \frac{1}{2}\left(C_R + C_L\right) \qquad\qquad a =
\frac{1}{2}\left(C_R - C_L\right)\ .
\end{equation*}} where the sum indicates that we can
couple the muon to any of the neutrinos, $\neutrino_i$.  The remaining
terms include the neutral fermion mass, $m_\internal$, and the vector
mass $M$.  In the limit that the model has no additional heavy neutral
fermionic states (that is, there are only the three light Standard
Model neutrinos), that those neutrinos are massless, and that the muon
mass is small compared to the vector mass, the above expression
reduces to
\begin{equation}
\amuon^\Wprime = \frac{m_\muon^2}{8\pi^2} \frac{\geff^2}{4
  M_\Wprime^2} \frac{10}{3} \left( C_L^2 +C_R^2 \right)\ .
\label{eq:charged}
\end{equation}
In particular, this result holds in the Standard Model case, where
$\geff = g$, $C_L = 1/\sqrt{2}$, and $C_R = 0$, giving a Standard
Model \Wpart\ contribution to \amuon\ of
\begin{equation*}
\amuon^\Wpart = \frac{m_\muon^2 \Gfermi}{8\pi^2 \sqrt{2}}
\frac{10}{3}\ ,
\end{equation*}
in agreement with the standard result.  

We can similarly derive an expression for the contribution of neutral
vectors at one-loop.  In Feynman gauge and applying the narrow width
approximation, we find
\begin{multline}
\amuon^\Zprime = -\frac{m_\muon g^2}{8\pi^2}
\int\limits_0^1 \dd{u} \frac{u(u-1)
  \left[ 2m_\muon (u-2)\left(vv^\dagger+aa^\dagger\right) +
    4m_\internal \left(vv^\dagger-aa^\dagger\right) \right]}{(1-u)M^2
  +um_\internal^2 +u(u-1)m_\muon^2}\\
+ \frac{m_\muon g^2}{8\pi^2} \frac{m_\internal}{M^2}
\int\limits_0^1 \dd{u} u^2
\frac{\left(m_\internal- m_\muon\right)^2 vv^\dagger
  -\left(m_\internal+ m_\muon\right)^2 aa^\dagger}{(1-u)M^2
  +um_\internal^2 +u(u-1)m_\muon^2}\\ 
- \frac{m_\muon g^2}{8\pi^2} \frac{m_\muon}{M^2}
\int\limits_0^1 \dd{u} u^2
\frac{\left\{\left(m_\internal- m_\muon\right)^2 vv^\dagger 
  +\left(m_\internal+ m_\muon\right)^2
  aa^\dagger\right\}(u-1)}{(1-u)M^2 +um_\internal^2
+u(u-1)m_\muon^2}\ ,
\end{multline}
where the second and third lines are the contributions from the
unphysical scalar diagram.  Again, most terms arise from the
Lagrangian couplings
\begin{equation*}
\mathcal{L} \sim \sum_i \geff \DB{\muon} \DG{\mu} \left(v^i + a^i
  \DG{5}\right) \lepton_i \Zprime_\mu + \text{h.c.} = \sum_i \geff
\DB{\muon}_L \DG{\nu} C_L^i \lepton_{Li} 
\Zprime_\nu + \geff \DB{\muon}_R \DG{\nu} C_R^i \lepton_{Ri}
\Zprime_\nu + \text{h.c.}\ .
\end{equation*}
We have explicitly included the possibility of flavor changing
neutral couplings; although most of the gauge extensions we will look
at contain a GIM-like mechanism that requires $\lepton_i = \muon$,
there are extensions where this is not the case.  When such tree level
flavor changing couplings are allowed, the $v$ and $a$ terms will
include the mixing factors, and we will have to sum over all possible
$\lepton_i$ that can circulate inside the loop.  In the limit where the
\muon\ and $\lepton_i$ masses are small compared to the vector mass,
the above expression reduces to
\begin{equation}
\amuon^\Zprime = -\frac{m_\muon}{8\pi^2}
\frac{\geff^2}{M_\Zprime^2} \frac{2}{3} \left(
  m_\muon\left(C_L^2 + C_R^2\right) - 3 m_\internal C_L C_R \right)\ ,
\label{eq:neutral}
\end{equation}
where we have explicitly retained the possibility that the new gauge
physics will admit flavor changing neutral current (FCNC) couplings.
In the Standard Model the GIM cancellation ensures that $m_\internal =
m_\muon$, and the gauge couplings are given by $\geff =
g/\cos\thetaW$, $C_L = -1/2 + \sin^2\thetaW$, and $C_R =
\sin^2\thetaW$.  The Standard Model \Znaught\ contribution to \amuon\ 
is then
\begin{equation*}
\amuon^\Znaught = -\frac{m_\muon^2 \Gfermi}{8\pi^2\sqrt{2}}
\frac{4}{3} \left( 1 +2\sin^2\thetaW -4\sin^4\thetaW \right)\ .
\end{equation*}
The contributions to \amuon\ from scalars with non-vanishing VEVs
(i.e. $v\neq0$) are negligible, and we will not consider this issue
further.  Detailed derivations of all of these expressions are
presented in \cite{companion} and compared to the results of the
references cited therein.

\section{Model Contributions to \amuon}
\label{sec:models}

In this section, we analyze the contributions to \amuon\ from a number
of explicit realizations of the three models presented at the
beginning of the previous section.  We divide this section into three
subsections, devoting one to each of the following classes of models:
the ununified (lepton-quark non-universal) models, the left-right
symmetric models, and the generation non-universal models.

\subsection{Ununified Models}
\label{sec:ununified}

In the Ununified Models, leptons and quarks are assigned charges under
different gauge groups.  The Ununified model of Georgi, Jenkins and
Simmons \cite{Georgi:1989ic,Georgi:1990xz} has the unbroken gauge
group
\begin{equation*}
\SUtwo_L^\ell \times \SUtwo_L^q \times \Uone_Y\ ,
\end{equation*}
where the left-handed leptons charged under $\SUtwo_L^\ell$ and the
left-handed quarks under $\SUtwo_L^q$.  The dominant additional
contributions to \amuon\ in this case are from the \Zprime\ and
\Wprime\ which couple to
\begin{equation*}
g \left(\frac{\cphi}{\sphi} T_q - \frac{\sphi}{\cphi} T_\ell
\right)\ ,
\end{equation*}
with $T_q = T_q^3$ for the \Zprime, $T_q = T_q^\pm$ for the \Wprime.
We have also used the shorthand $\sphi = \sin\phi$ and $\cphi =
\cos\phi$.  In this case, the contributions to \amuon\ can be
determined from Equations~\ref{eq:charged} and~\ref{eq:neutral}; we
find
\begin{equation}
\amuon^{\text{UUM}} = \frac{m_\muon^2 \Gfermi}{8\pi^2 \sqrt{2}}
2 \frac{\sphi^2}{\cphi^2} \frac{M_\Wpart^2}{M_\Wprime^2}\ ,
\end{equation}
where we have combined the \Zprime\ and \Wprime\ contributions, since
$M_\Zprime = M_\Wprime$ to lowest order.  Limits obtained on this
Ununified model from precision electroweak data
\cite{Chivukula:1995qw} can be used to find upper bounds on this
contribution.  The largest possible value of this contribution is less
than $10^{-11}$ (except for very large $\sphi$, which is precisely the
region where our approximations break down and our calculations no
longer apply due to the large corrections to the \Znaught\ and \Wpart\ 
couplings compared to the Standard Model), and hence these vectors can
not by themselves explain the observed discrepancy.

We could also consider an ununified model with gauge group
\begin{equation*}
\SUtwo_L \times \Uone_Y^\ell \times \Uone_Y^q\ ,
\end{equation*}
with a \Zprime\ coupling
\begin{equation*}
g' \left( \frac{\cphi}{\sphi} Y_q - \frac{\sphi}{\cphi} Y_\ell \right)
\end{equation*}
and of course there is no \Wprime.  The contribution to \amuon\ is
then given by
\begin{equation}
\amuon^\Zprime = \frac{m_\muon^2 \Gfermi}{8\pi^2 \sqrt{2}}
\frac{4}{3} \frac{M_\Znaught^2}{M_\Zprime^2} \frac{\sphi^2}{\cphi^2}
\sW^2\ .
\end{equation}
We can use the experimental value of \amuon\ to place a limit on the
value of $M_\Zprime$ necessary to fully account for the observed
discrepancy 
\begin{equation*}
\frac{M_\Zprime}{\tphi} < \unit[26]{GeV}\ .
\end{equation*} 
In order not to disagree with the LEP precision observables, however,
\sphi\ must be small (otherwise the contribution to leptonic
observables near the \Znaught\ pole from the \Zprime\ coupling above
would be large), hence \tphi\ must be smaller than 1.  This
requirement effectively rules out a model of this type as the sole
explanation for the \amuon\ discrepancy, as such a light \Zprime\ 
would have been observed at CERN LEP and the Fermilab Tevatron.

\subsection{Left-Right Models}
\label{sec:lr}

In the Left-Right Models, left- and right-handed fermion doublets
transform under different gauge groups.  While there are many ways to
build such models (for a brief overview and references, see
\cite{Groom:2000in}), we choose to analyze a ``generic'' model with
the gauge group
\begin{equation*}
\SUtwo_L \times \SUtwo_R \times \Uone_{(B-L)}\ ,
\end{equation*}
where the left-handed fermions transform as doublets under $\SUtwo_L$,
the right-handed fermions transform as doublets under the $\SUtwo_R$,
and both left- and right-handed fields charged under the $B-L$
hypercharge (baryon number minus lepton number).  The coupling of the
heavy neutral gauge field, \Zright, is given by
\begin{equation*}
g' \left(\frac{\cphi}{\sphi} \frac{1}{2}\left(B-L\right) -
  \frac{\sphi}{\cphi} T_R^3\right)\ ,
\end{equation*}
while the coupling to the heavy charged field, \Wright, is given by
\begin{equation*}
\frac{g'}{\cphi} T_R^\pm\ .
\end{equation*}
We can determine \amuon\ from Equation~\ref{eq:neutral}, and we obtain
\begin{equation}
\amuon = \frac{m_\muon^2 \Gfermi}{8\pi^2 \sqrt{2}}
\frac{M_\Znaught^2}{M_\Zright^2} \sW^2 \left( \frac{10}{3}
  \frac{1}{\sphi^2\cphi^2} - \frac{4}{3} \left(
    \frac{\sphi^2}{\cphi^2} - \frac{\cphi^2}{\sphi^2} + 1 \right)
\right)\ .
\end{equation}
Assuming a very conservative \unit[700]{GeV} lower bound on the
mass\footnote{Bounds on $M_\Zright$ are usually quoted as bounds on
  parameters which are equivalent to placing bounds in the
  $M_\Zright$-$\phi$ plane, not as simultaneous bounds on $M_\Zright$
  and $\phi$ as for the other models discussed in this paper.  See
  references in \cite{Groom:2000in}.  It is thus not possible to
  provide limits on these models in the same form as for the other
  models considered in this paper.}  of the \Zright, we find that the
\Zright/\Wright\ contribution to \amuon\ is less than
$35\times10^{-11}$ for $0.25 < \sphi < 0.99$.  For larger or smaller
values of \sphi, the assumptions we made in deriving our results break
down; in these extremes, the contributions to \amuon\ will be
dominated by the shifts in electroweak couplings, and not from the new
gauge fields.  These are precisely the regions of parameter space that
should be ruled out by the electroweak data.  Hence, the discrepancy
can not be entirely explained by this type of model.

\subsection{Generation Non-universal Models}
\label{sec:generational}

In the Generation Non-universal models, the third generation fermions
are charged under a different gauge group than the first and second
generation fermions.  We will consider two models.

Generation non-universal models arise in certain extended technicolor
models \cite{Chivukula:1994mn,Chivukula:1996gu}, and the topflavor
models \cite{Malkawi:1996fs}.  Here we examine the non-commuting
extended technicolor (NCETC) scenario due to Chivukula, Simmons, and
Terning \cite{Chivukula:1996gu}.  This model has the gauge group
\begin{equation*}
\SUtwo_L^\ell \times \SUtwo_L^h \times \Uone_Y\ ,
\end{equation*}
where $\SUtwo_L^\ell$ couples to the left-handed first and second
generation fermions (the $\ell$ight generations) and $\SUtwo_L^h$
couples to the left-handed third generation fermions (the $h$eavy
generation).  The coupling is given by
\begin{equation*}
g \left(\frac{\cphi}{\sphi} T_h - \frac{\sphi}{\cphi} T_\ell \right)\
, 
\end{equation*}
where again, the $T_i = T_3 \; (T^\pm)$ for the $\Zprime\;
(\Wprime)$.\footnote{Our notation for \sphi\ and \cphi\ are the
  opposite of those used in \cite{Chivukula:1996gu}.} Again, we can
determine the contribution to \amuon\ from the results in
Equations~\ref{eq:charged} and~\ref{eq:neutral}, and we obtain
\begin{equation}
\amuon^{\text{NCETC}} = \frac{m_\muon^2 \Gfermi}{8\pi^2 \sqrt{2}}
2 \frac{\sphi^2}{\cphi^2} \frac{M_\Wpart^2}{M_\Wprime^2}\ . 
\end{equation}
The constraints from precision electroweak data on the ``light'' case
of NCETC in \cite{Chivukula:1996gu} imply that the largest possible
contribution to \amuon\ is smaller than $\unit{3\times10^{-11}}$.
Thus, this extension alone can not explain the discrepancy.

There are other generation non-universal models; Topcolor Assisted
Technicolor (TC2) \cite{Hill:1995hp}, for example, contains an extend
weak sector with gauge group
\begin{equation*}
\SUtwo_L \times \Uone_Y^\ell \times \Uone_Y^h\ .
\end{equation*}
The coupling of the \Zprime\ is given by
\begin{equation*}
g'\left( \frac{\cphi}{\sphi} Y_h - \frac{\sphi}{\cphi} Y_\ell \right)
\end{equation*}
If we choose to assign fermionic charge for the muon under
$\Uone_Y^\ell$ as in the Standard Model, then scaling from the
Standard Model \Znaught\ contribution, we find
\begin{equation}
\amuon^\Zprime = \frac{m_\muon^2 \Gfermi}{8\pi^2 \sqrt{2}}
\frac{4}{3} \frac{M_\Znaught^2}{M_\Zprime^2} \frac{\sphi^2}{\cphi^2}
\sW^2\ .
\end{equation}
Chivukula and Terning have used precision electroweak data to
constrain the parameters of this TC2 model \cite{Chivukula:1996cc}
(our hypercharge assignment is their ``optimal'' scenario, which we
label OTC2); using their results, we find that the OTC2 contribution
to \amuon\ can be no greater than about $\unit{0.3\times10^{-11}}$.
Hence, this model can not explain the discrepancy by itself.

In a gauge theory with a larger gauge group than the Standard Model
where the couplings of the fermions are not generation universal,
there will arise, in the absence of additional symmetries, tree level
mixings between fermion mass eigenstates at gauge-fermion-fermion
vertices, even if all neutrino masses are
zero.\cite{Huang:2001zx,Langacker:2000ju,Malkawi:1996fs,Rador:1998is}
In other words, there will be no automatic GIM cancellation in the
extended neutral current interactions, although the SM neutral
currents will still admit an approximate GIM mechanism in these cases.
If we don't eliminate these couplings (with additional discrete flavor
symmetries, for example), we have to consider the possibility that
heavier fermions may propagate on the internal lines of the \Zprime\ 
diagram.  From Equation~\ref{eq:neutral}, we see that heavy internal
fermions can make potentially large contributions to \amuon.  Let us
see how this works.

Consider the extended neutral current Lagrangian with the fermions in
the ``gauge basis'' (where the Lagrangian is diagonal in the gauge
basis flavor space, but where the coupling matrix is not necessarily a
multiple of the identity)
\begin{equation*}
\mathcal{L} = \PsiGbar i \DS{D} \PsiG = \PsiGbar i \DG{mu} \left\{
D_\mu^{\text{SM}} -i g_\Zprime \Zprime_\mu C_\Zprime 
\right\} \PsiG\ ,
\end{equation*}
where \PsiG\ is a vector of charged fermions, and $C_\Zprime$ is the
vertex operator matrix in the gauge basis.  It is important to note
that $C_\Zprime$ is diagonal, but is not a multiple of the identity:
$C_\Zprime \neq \alpha I$.  We now perform a rotation to the gauge
basis $\PsiG = \Lambda_{GM} \PsiM$.  Inserting this rotation, we find
the Lagrangian in the fermion mass basis
\begin{equation*}
\mathcal{L} = \PsiMbar i \DG{\mu} \left\{ D_\mu^{\text{SM}} -i
  g_\Zprime \Zprime_\mu L_\Zprime \right\} \PsiM\ ,
\end{equation*}
where $L_\Zprime = \Lambda_{GM}^\dagger C_\Zprime \Lambda_{GM}$, which
may not even be diagonal, permitting tree level flavor changing
couplings in the extended neutral current sector.  

What couplings do we find for these flavor changing interactions?
Consider the diagonal (flavor conserving) elements in a three
generation model, for example the $L_\Zprime^{\tauon\tauon}$ element
\begin{equation*}
L_\Zprime^{\tauon\tauon} = \sum\limits_{G=1}^{3}
\Lambda_{G\tauon}^\dagger C_\Zprime^{GG} \Lambda_{G\tauon}\ .
\end{equation*}
Assume that $C_\Zprime^{11} = C_\Zprime^{22} \neq C_\Zprime^{33}$.
Now, applying three-generation unitarity and rearranging, we find
\begin{equation*}
L_\Zprime^{\tauon\tauon} = C_\Zprime^{11} + \left(C_\Zprime^{33} -
  C_\Zprime^{11} \right) \Lambda_{3\tauon}^\dagger \Lambda_{3\tauon}\
.
\end{equation*}
In the limit of small off-diagonal mixing, this expression
  simplifies to
\begin{equation*}
L_\Zprime^{\tauon\tauon} \approx C_\Zprime^{33}\ ,
\end{equation*}
as we might have assumed.  For the off-diagonal (flavor violating)
terms, for example the $L_\Zprime^{\tauon\muon}$ coupling, we find
\begin{equation*}
L_\Zprime^{\tauon\muon} = \sum\limits_{G=1}^3
\Lambda_{G\tauon}^\dagger C_\Zprime^{GG} \Lambda_{G\muon}\ .
\end{equation*}
Applying three-generation unitarity and rearranging, we find
\begin{equation*}
L_\Zprime^{\tauon\muon} = \left(C_\Zprime^{33} - C_\Zprime^{11}\right)
\Lambda_{3\tauon}^\dagger \Lambda_{3\muon}\ .
\end{equation*}

We can now consider toy flavor mixing extensions to the generation
non-universal models studied above (NCETC and OTC2).  We assume these
toy models have the following properties:
\begin{enumerate}
\item There are only three generations of fermions.
\item The left- and right-handed flavor rotations that diagonalize the
  fermion mass matrix are the same; this is certainly not required,
  but greatly simplifies the calculations.
\item Since constraints on processes such as $\muon \to
  \electron\photon$ and $\muon \to \electron\electron\electron$ are
  rather stringent, we assume that there are no
  $\electron\muon\Zprime$ or $\electron\tauon\Zprime$ vertices; that
  is, only the \tauon\ and \muon\ mix.
\item The $\muon\tauon$ mixing is small.
\end{enumerate}
With these assumptions, we can calculate the additional contributions
using Equations~\ref{eq:charged} and~\ref{eq:neutral}.  For NCETC, we
find that \amuon\ is given by
\begin{multline}
\amuon^{\text{NCETC}} \approx \frac{m_\muon^2 \Gfermi}{8\pi^2 \sqrt{2}}
\frac{10}{3} \frac{\sphi^2}{\cphi^2} \frac{M_\Wpart^2}{M_\Wprime^2} - 
\frac{m_\muon^2 \Gfermi}{8\pi^2 \sqrt{2}}
\frac{4}{3} \frac{\sphi^2}{\cphi^2} \frac{M_\Wpart^2}{M_\Wprime^2}
\left|\Lambda_{2\muon}^\dagger \Lambda_{2\muon} \right|^2 -\\
\frac{m_\muon^2 \Gfermi}{8\pi^2 \sqrt{2}}
\frac{4}{3} \left(\frac{\cphi}{\sphi} - \frac{\sphi}{\cphi}\right)^2
\frac{M_\Wpart^2}{M_\Wprime^2} \left|\Lambda_{3\tauon}^\dagger
  \Lambda_{3\muon} \right|^2\ .
\end{multline}
Lacking experimental data on these mixing matrices, we will have to
make some assumptions to obtain numerical predictions.  If we take,
for example, $\left|\Lambda_{3\tauon}\right|^2 =
\left|\Lambda_{2\muon}\right|^2 = 1 - \left|\Lambda_{3\muon}\right|^2
= 1 - \left|\Lambda_{2\tauon}\right|^2 = 0.99$,
$\left|\Lambda_{1\electron}\right| = 1$, and all others zero, we find
a limiting contribution that is almost unchanged from the no-mixing
case, with contributions less than $3\times10^{-11}$ over the whole
parameter space.  The result is unchanged because there are no
right-handed couplings to the \Zprime, and hence no possibility of
enhancement from the larger internal fermion (see
Equation~\ref{eq:neutral}).

For the OTC2 model, we find that the dominant new contributions are
further enhanced by $m_\tauon/m_\muon$ compared to the NCETC values,
plus potential gauge group mixing angle enhancements,
\begin{equation}
\amuon^{\text{OTC2}} = \frac{m_\muon^2 \Gfermi}{8\pi^2 \sqrt{2}}
\frac{4}{3} \frac{M_\Znaught^2}{M_\Zprime^2} \frac{\sphi^2}{\cphi^2}
\sW^2 \left|\Lambda_{2\muon}^\dagger \Lambda_{2\muon} \right|^2 + 
\frac{m_\muon m_\tauon \Gfermi}{8\pi^2\sqrt{2}} 8
\frac{M_\Znaught^2}{M_\Zprime^2}  \left(\frac{\cphi}{\sphi} -
  \frac{\sphi}{\cphi}\right)^2 \sW^2 \left|\Lambda_{3\tauon}^\dagger
  \Lambda_{3\muon} \right|^2\ .
\end{equation}
Using the same mixing angle parameters as above, we find that the
mixing angle enhancement at small \sphi\ overlaps with a small window
of low \Zprime\ mass in the precision data, allowing contributions of
up to $35\times10^{-11}$; however, over most of the mixing angle
parameter space the contributions are two orders of magnitude smaller.
This large enhancement is due the existence of right-handed couplings
which result in a large $m_\tauon/m_\muon$ enhancement when the tau
circulates in the loop.  It is obviously possible to obtain even
larger enhancements if there are additional heavy fermionic states
with the appropriate quantum numbers to mix with the muon at the
\Zprime\ vertex.

\section{Conclusion}
\label{sec:conclusion}

We have presented general expressions for the contributions of neutral
and charged vector bosons to the anomalous magnetic moment of the
muon.  We then looked at a number of simple, but representative,
models with extended electroweak gauge structures, and calculated
their contributions to \amuon.  We found that, in general, models with
different gauge interactions for the leptons and quarks (the ununified
models), and models with different gauge interactions for the heavy
and light fermions (the generation non-universal models) with flavor
diagonal couplings, are both constrained by precision electroweak data
and can make only very small contributions to \amuon, of order
$10^{-11}$.  Models with simple extended left-right symmetries can
generally provide only small contributions to \amuon, of order
$10^{-11}$, in those regions of parameter space where they are not
expected to disturb the precision electroweak data.  Interestingly,
the generation non-universal models, as they admit the possibility of
flavor changing tree level couplings, can provide potentially large
contributions to \amuon; even with very small mixing between the muon
and the tau, some of these models can generate contributions of up to
$35 \times 10^{-11}$.  If such charged sector mixing is observed, one
might be able to indirectly observe the new gauge bosons at the next
generation of electron-positron and hadron collider experiments.

However, of the models we have examined, none can fully explain the
observed discrepancy between the current experimental value of \amuon\ 
and the Standard Model prediction, even if we assume that the masses
of the new gauge bosons are as small as those allowed by precision
electroweak data.  In the context of models with additional matter
fields as well as the additional gauge fields discussed here, we have
shown that some of the gauge field contributions to \amuon\ can be
substantial and important, in particular those involving non-universal
lepton couplings.  The final results of the E821 Collaboration along
with the results of high energy collider experiments will soon be able
to tell us much more about the possible existence and properties of
these extended electroweak interactions.

\vspace{24pt}
\begin{center}\textbf{Acknowledgments}\end{center} 
\vspace{12pt} 

I thank E.~H.~Simmons and T.~Rador for insightful discussions and
comments on the manuscript. This work was supported in part by the
Department of Energy under grant DE-FG02-91ER40676, the National
Science Foundation under grant PHY-0074274, and by the Radcliffe
Institute for Advanced Study.

\bibliographystyle{apsrev}
\bibliography{gauge}

\end{document}